# Mal-Netminer: Malware Classification Approach Based on Social Network Analysis of System Call Graph

Jae-wook Jang,[1] Jiyoung Woo,[1] Aziz Mohaisen,[2] Jaesung Yun,[1] and Huy Kang Kim[1]

[1]*Graduate School of Information Security, Korea University, Seoul 136-713, Republic of Korea*
[2]*Computer Science and Engineering Department, State University of New York at Buffalo (SUNY Buffalo), Buffalo, NY 14260-2500, USA*

As the security landscape evolves over time, where thousands of species of malicious codes are seen every day, antivirus vendors strive to detect and classify malware families for efficient and effective responses against malware campaigns. To enrich this effort and by capitalizing on ideas from the social network analysis domain, we build a tool that can help classify malware families using features driven from the graph structure of their system calls. To achieve that, we first construct a system call graph that consists of system calls found in the execution of the individual malware families. To explore distinguishing features of various malware species, we study social network properties as applied to the call graph, including the degree distribution, degree centrality, average distance, clustering coefficient, network density, and component ratio. We utilize features driven from those properties to build a classifier for malware families. Our experimental results show that "influence-based" graph metrics such as the degree centrality are effective for classifying malware, whereas the general structural metrics of malware are less effective for classifying malware. Our experiments demonstrate that the proposed system performs well in detecting and classifying malware families within each malware class with accuracy greater than 96%.

## 1. Introduction

Despite the increasing efforts and investments antivirus (AV) vendors are making to defend against the spread of malware families, malware infection is still one of today's most serious threats in the global security landscape. Malware infection is considered the first step in many attacks launched by cyber criminals, and the ever increasing numbers of malware families have made defense against those criminals a difficult task. According to recent reports by AV-TEST [1], approximately 60 million new pieces of malware are reported for the period from January 2013 to December 2014. Techniques utilized for creating those malware pieces have evolved over time, and malware authors create new malware variants employing various circumvention techniques, such as encryption, polymorphism, and obfuscation. To defend against malware, AV vendors analyze tens of thousands of pieces of malware every day and prevent them from spreading, thus putting themselves and cyber criminals in an endless arms race.

Cyber criminals can easily create malware variants with the same semantics by reusing the same core code. Although they generate many malware variants for the same malware family, the base malware families have the invariant characteristics and patterns, of malicious behaviors. Those invariant characteristics can be derived from the instruction or binary code of the malware. Utilizing signature-based techniques to capture the similarity between the base family and its variants has several shortcomings. Using polymorphism or metamorphism techniques, malware can evade the detection technique while maintaining its behavior unchanged.

Cyber defenders, including AV vendors, are not reactive to malware and generate signatures to partly or wholly address the obfuscation and encryption circumvention techniques [2]. However, signature-based methods require human intervention to construct signatures based on domain knowledge, and defenders should update the signature databases with new signatures continuously. While these approaches are effective for known malware, they cannot detect unknown malware, particularly zero-day attacks. To overcome those shortcomings, the research community has established the alternative of behavior-based methods for malware detection utilizing dynamic analysis of malicious binaries.

Using dynamic analysis, various prior studies proposed malware analysis methods that exploit one or more behavioral aspects of the malware execution, including statistical methods leveraging the system or API call sets [3–5], instruction pattern sets [6], or call graph matching [7, 8]. The prior work assumes that the malicious behavior of a malware sample is determined by a distinguishing call or instruction set. Since such call or instruction sets form a specific frequency distribution or have unique call sequences, they can be used as one of the metrics for classifying malware. However, these methods are weak in detecting malware variants generated by polymorphism techniques (e.g., statement reordering or junk code insertion) and can hardly detect new malware or variants. The prior call graph-based methods are mainly based on similarity search [7, 8]. Furthermore, the similarity matching algorithms utilized to achieve this end goal are computationally heavy and can result in high false alarms. The computational overhead associated with the similarity matching makes it hard to produce a real-time system for detection and classification—a feature often desirable in many industrial systems. In addition, these methods using graph properties are limited to depicting the structural information of malware itself.

To overcome drawbacks of previous system call-based techniques, we propose a novel and lightweight classification method that examines the topological structure and influence properties of system call graph—in a way analogous to graph properties in the social network analysis. We assume that the malicious behavior is characterized by differences in the system calls of a malware. As with social networks analysis, the system call set has an influence on the behavior of the program. Therefore, we can detect the malicious behavior of programs by generating the system call graph of the malware and exploring its network properties. We prepare a malicious system call dictionary to classify malware families within each malware class (in this paper, "malware class" denotes malware type (e.g., adware, Trojan, and worm)). As we examine various network properties of system calls found in malware, we can classify malware families within each malware class.

*Contribution.* It is as follows:

(i) We propose a novel classification method that exploits network properties that are heavily studied in the social network analysis field, such as the degree distribution, degree centrality, and average distance. To the best of our knowledge, this work is the first attempt to examine the topological and node properties of malicious system calls in a call graph. Utilizing a small feature set for classification, our system is scalable and lightweight, compared to other methods from the literature. In practice, the proposed system enables AV vendors to react to malware in a scalable and lightweight manner. In building the main system call graph used for extracting the features for classification, our method only requires statistical values related to the network analysis function and does not impose any additional computational overhead.

(ii) We demonstrate experimentally and using real-world data the capabilities of our system in detecting various malware families with accuracy greater than 96%.

(iii) While our method mainly utilizes dynamic analysis for malware classification, it is also flexible in utilizing static analysis by extracting API call features.

*Organization.* The rest of this paper is organized as follows. In Section 2, we review the previous work. In Section 3, we review preliminaries of the social network analysis. Data exploration to find meaningful features for automatic classification is presented in Section 4. In Section 5, we present our classification method. In Section 6, we present the results and performance evaluation. In Section 7, we discuss the limitation of our proposed method. Finally, in Section 8, we state our conclusions and future research directions.

## 2. Related Work

Most of the previous work on detection or classification of malware used signature-based or behavior-based methods. Table 1 summarizes such methods in the literature. In the following, we review some of this literature in more details.

*2.1. Static Analysis.* Static analysis techniques rely on examining the binary code to determine its properties without actually executing it. There are two types of methods of static analysis depending on the features utilized for operation: statistical and graph matching-based methods. In the statistical method, defenders transform the binary code of malware into an assembly code to extract and analyze characteristics, such as the $n$-grams of instruction or call patterns. The graph matching method is mainly based on similarity matching. For that, defenders build call graphs (e.g., system call graph, function-call graph, or API call graph), compare graphs with each other, and classify malware based on how well defenders match with previously known behaviors of the given malware species. Using polymorphism or metamorphism techniques, malware can disguise its appearance while keeping its behavior unchanged. As a result, such techniques using signature-based methods are easily evaded by cyber criminals.

Researchers also looked at realizing static analysis on instruction or call pattern. Christodorescu and Jha [13] implemented a static analyzer named SAFE, which can analyze executables using predefined malicious instruction patterns. Kolter and Maloof [14] demonstrated how text classification can be applied to malware classification by extracting byte sequences from executables, converting those byte sequences into $n$-grams and constructing several classifiers. Reddy and Pujari [15] proposed a detection method utilizing the frequency pattern of the $n$-grams in binaries as an alternative method to measure information gain. Tabish et al. [16] proposed a malware detection method based on the analysis of byte-level file content. They computed a wide range of statistical and information-theoretic features in a block-wise manner to quantify the byte-level file content. Shankarapani et al. [17] have proposed the malware detection methods named SAFE and MEDiC; SAFE used API call sequences and MEDiC used assembly codes as feature vectors. Sung et al. [18] proposed the signature-based method named SAVE, and they compared the API sequences of malicious files with those of a signature database. Wang et al. [19] proposed a malware detection method that can identify suspicious behavior through the frequency of a predefined API call set.

Table 1: Various static and dynamic analysis approaches in previous works.

| Approach | Method | Viewpoint | Previous works |
|---|---|---|---|
| Static analysis | Graph matching | Instruction | [9] |
| | | API call | [10–12] |
| | Statistical method (frequency and sequence) | Instruction | [13] |
| | | Bytecode | [14–16] |
| | | API call | [17–19] |
| Dynamic analysis | Graph matching | Instruction | [7] |
| | | System call | [8, 20] |
| | Statistical method (frequency and sequence) | Instruction | [6] |
| | | System call | [3, 5] |
| | | API call | [4] |

Many researchers used call graphs as structural information in detecting malware. Bruschi et al. [9] proposed a self-mutating malware detection method that compares the control flow graph of a target program with that of known malware. Hu et al. [10] proposed a malware database management system named SMIT, which can efficiently make determination based on the function-call graph of malware. SMIT also used properties of the function-call graph to implement an efficient nearest-neighbor search algorithm for large malware datasets. Lee et al. [11] proposed a metamorphic malware detection method. They converted API call sequences into a call graph to extract the semantics of the malware and reduced the call graph to a code graph for computational cost reduction. Wüchner et al. [12] proposed a behavioral malware detection approach based on a system-wide quantitative data flow model. To gather the relevant information, they leveraged API calls.

*2.2. Dynamic Analysis.* Dynamic analysis approaches run executables inside the isolated environment to capture the runtime behavior. These approaches extract behavioral characteristics through taint analysis on the relation of system or API caller-callee. This approach addresses obfuscation, packing attempts, and self-modification, since all of those approaches are eliminated during the execution of malware [21].

Many methods are proposed for dynamic analysis using instruction or call patterns. Anderson et al. [7] proposed a malware detection method that analyzes graphs constructed from instruction traces. They extended the 2-gram method utilizing the transition probabilities of a Markov chain; they treated the Markov chain as a graph and used the graph kernel to construct a similarity matrix, which is then used for analyzing malware samples based on their similarity. Fredrikson et al. [20] introduced an automatic malware detection system (Holmes), which extracts dependence graph with system calls, mines significant behaviors from samples, and synthesizes an optimally discriminative specifications from mined behaviors. Kolbitsch et al. [8] proposed a malware detection method utilizing system call patterns based on data flow dependencies. They tracked dependencies between system calls and compared the runtime behavior of unknown programs with pregenerated behavioral graphs.

Dai et al. [6] proposed a malware detection method using dynamic instruction sequences. They converted runtime instruction sequences into basic blocks and extracted frequent instruction groups for instruction's relative frequency calculation as features for a classification model. Bayer et al. [3] proposed a malware clustering architecture using taint analysis with system calls. They extracted behavioral profiles by abstracting the system calls, their dependencies, and network activities and applied locality sensitive hashing on the behavioral profiles to achieve efficient and scalable malware clustering. Yuxin et al. [5] proposed a malware detection method that analyzes binary code to derive system call sequences and utilizes features based on $n$-grams. Bayer et al. [4] introduced TTAnalyze, a system for analyzing the behavior of an unknown program by executing the code. Their tool records security-relevant system or API calls triggered in the execution and generates reports including file activity, registry activity, and network activity.

## 3. Background

When a program is executed, it invokes a series of relevant system calls. For example, in order to open a specific file and then write to it, a program successively invokes *NtCreateFile, NtCreateSection, NtMapViewofSection*, and *NtWriteFile*. The intermediate system calls, namely, the *NtCreateSection* and *NtMapViewofSection*, play a role in bridging between source and destination system calls. In other words, system calls interact with other system calls, whereas intermediate system calls pass information to destination system calls.

In an analogous manner, we can explain this behavior through the lenses of social network analysis techniques as a tool. Social network analysis (SNA) has its origin in social science and scientific theories such as network analysis and graph theory. Network analysis concerns itself with the foundation and problem-solving pertaining to transforming problem into a network structure in the form of graph and using well-established algorithms for understanding such graphs.

Graph theory provides a set of the abstract concepts and methods for the analysis of such graphs. SNA in social science conceptualizes social structure as a network with ties connecting members. SNA can also conceptualize a range of problems in other domains, such as social networks, technological networks, and information networks. Technological networks, man-made networks designed for distribution of commodity or resource, such as electricity or information, are also studied using the same tools. The network of mobile stations is a good example for such applications, where mobile network operators use SNA to optimize the structure and capacity of their networks [22]. Other technological networks that have been studied include the network of electric power grid, airline routes, roads networks, railways, and pedestrian traffic networks [23].

In a similar way, when malware is executed it consecutively calls out a predefined set of system calls. We record those system calls as a sequence and construct a system call graph from this sequence. The system call graph is a directed graph $G = (V, E)$, where $V$ is a set of nodes and $E$ is a set of edges. A node represents a system call that is invoked, and an edge is determined by the call sequence of the system calls in $V$: $E = \{\langle v_i, v_j \rangle \mid v_i, v_j \in V, v_i \in V_m \text{ or } v_j \in V_m\}$, where $v_i$ denotes the former system call in the sequence and $v_j$ denotes the system call right after it. $V_m$ represents element of malicious system call dictionary, $V_m \subseteq V$. We now define the system call graph more formally.

*Definition 1* (system call graph). A system call graph is a directed graph defined by 4 tuples as follows: $g = (V_g, E_g, L_g, l_g)$ where $V_g$ is the vertex set representing the system calls, $E_g \subseteq V_g \times V_g$ is the edge set corresponding to a function, $L_g$ is the set of labels that can identify each system call, and $l_g$ is the labeling function that assigns labels to vertices.

We transform such call sequence into a system call graph to examine the social network properties through SNA tools. We extract system call trace from a target program as illustrated in Figure 1(a). Analyzing dependency among system calls, we build a system call sequence (or a system call graph) as shown in Figure 1(b) [24]. In particular, as tracking the arguments of system call (e.g., *FileHandle* and *SectionHandle*), we know dependency or sequence among system calls and build a system call graph of the target program.

By comparing nodes of call graph with predefined malicious system call dictionary, we extract system calls found in malware and their one-hop neighbor (refined system call sequence); we finally make such refined system call sequence to depict a system call graph as illustrated in Figure 1(b). The system call graph we defined consists of nodes included in the predefined malicious system call dictionary and nodes that are one-hop neighbors of aforementioned nodes. For example, in Figure 1(b), one-hop neighbor of *NtCreateSection* is *NtCreateFile* and *NtMapViewofSection*. Since *NtCreateSection* exists in the malicious system call dictionary, *NtCreateSection* establishes edge with *NtCreateFile* and *NtMapViewofSection*. We disallow the presence of multiple edges but allow the presence of self-edge (loops), when building the system call graph. To measure the influence of system calls found in malware, we generate a system call graph and adopt both the centrality and cohesion measures for understanding the underlying graph. Centrality is a measure of how many connections one node has with others and captures the prominence of a node in the network. Centrality measures include the degree, betweenness, and closeness centralities. On the other hand, cohesion is a measure to characterize the structure of a network. Cohesion measures include the clustering coefficient, average distance, network density, and component ratio; cohesion measures are quantified as a conjunct relation among nodes in the perspective of the network level in a system call graph.

In the following, we define several measures of centrality and cohesion and use them to explore the previously defined graph for distinctive features of malware families' identification and classification.

*3.1. Degree Distribution.* The degree distribution tells us the network structure and is one of the most fundamental network properties [25–27]. The degree distribution is the probability distribution of degrees over the whole network. The degree distribution $p(k)$ is defined as the fraction of nodes with degree $k$ in the network. If there are $n$ nodes in the network and $n_k$ nodes with degree $k$, we define $p(k) = n_k/n$.

A large majority of nodes have low degree, whereas a small number of nodes have high degree; such nodes are known as hubs. A minority of nodes with high degree have more influence. When the degree distribution on logarithmic scales is approximately represented by a straight line, we say that the distribution follows a power-law distribution [25, 26].

Newman [25] proved that many real networks, such as the Internet topology, networks for city population, word frequency, earthquakes magnitude, and moon creator diameter, have degree distributions with a tail of high degree.

*3.2. Degree Centrality.* The degree centrality is a measure of how well a node is connected with other nodes in terms of directed connections. This is, the degree centrality is the number of links that a node has with others [26, 28]. A directed graph has in- and out-degree centralities, defined as follows:

(1) *In-degree*: defined as the number of ties directed toward the node.

(1) Process 2160 starting at 00402264
(2) C:₩Users₩···₩Sample.exe
(3) Loaded DLL at 77250000 ntdll.dll
(4) ···
(5) *NtCreateFile*(*FileHandle* = A, ..., ObjectAttributes ="Sample.exe")
(6) ···
(7) *NtCreateFile*(*FileHandle* = B, ..., ObjectAttributes = "1111.exe")
(8) ···
(9) *NtCreateSection*(*SectionHandle* = C, ..., *FileHandle* = B)
(10) *NtMapViewOfSection*(*SectionHandle* = C, ···)
(11) *NtWriteFile*(*FileHandle* = B, ···)
(12) ···
(13) *NtQueryInformationFile*(*FileHandle* = A, IoStatusBlock = ···)
(14) ···

(a) System call trace

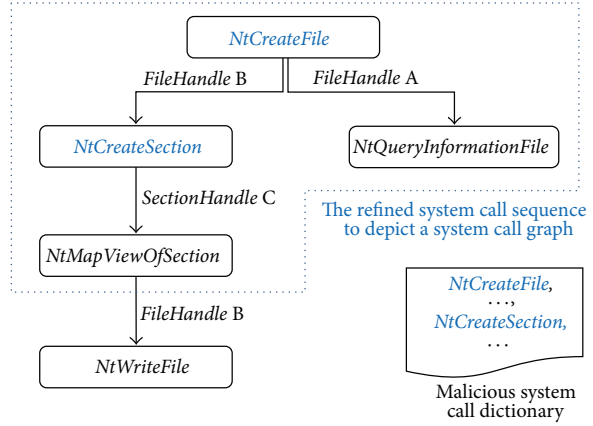

(b) A corresponding call graph

FIGURE 1: Procedure of a converting system call trace into a call graph.

(2) *Out-degree*: defined as the number of ties that the node directs toward others.

With the in-degree centrality, a node that has many ties is characterized as prominent node. On the other hand, with the out-degree centrality, a node that has a higher out-degree centrality is characterized as an influential node.

### 3.3. Betweenness Centrality.
The betweenness centrality is a measure of the degree to which a node serves as a bridge on the route between other nodes in the graph [26, 28]. The between centrality is often used as an index of the potential of a node for controlling communication between other nodes. The betweenness centrality of a node $i$, $X_i$ is defined as

$$X_i = \sum_{s,t} \frac{n_{st}^i}{g_{st}}, \qquad (1)$$

where $n_{st}^i$ denotes the number of geodesic paths from node $s$ to node $t$ which pass through node $i$ and $g_{st}$ denotes the total number of geodesic paths from node $s$ to node $t$.

### 3.4. Clustering Coefficient.
The clustering coefficient measures the average probability in which two neighbors of a node are neighbors [26, 28] and measures the density of triangles in the network. We define the clustering coefficient as the fraction of transitive triples or the fraction of closed paths of length two in the network.

### 3.5. Average Distance.
The distance is the ratio of the number of edges in the geodesic path between all pairs of nodes to the number of the geodesic path between all pairs of nodes in the graph. Given a connected network, the average distance (AD) is given by

$$\text{AD} = \frac{\sum_{\{i,j\}: l_g(i,j) \neq \infty} l_g(i,j)}{\left|\{\{i,j\} : l_g(i,j) \neq \infty\}\right|}, \qquad (2)$$

where $l_g(i,j)$ denotes the number of edges in the geodesic path connecting nodes $i$ and $j$ in the network.

### 3.6. Network Density.
The network density is the ratio of the number of edges in the network to the total number of possible edges between all pairs of nodes. That is, the network density is defined as the fraction of edges that are actually present in the graph [26, 28]. Assuming no multiedge or self-edge in the graph, the network density is given by

$$\rho = \frac{m}{\binom{n}{2}} = \frac{2m}{n(n-1)} = \frac{k}{n-1}, \qquad (3)$$

where $n$, $m$, and $k$ denote the number of nodes, the number of edges, and the average degree, respectively. As the number of nodes becomes infinite, the network density becomes $k/n$ and the mean degree becomes a constant.

### 3.7. Component Ratio.
A component is a subset of the nodes of a network such that there exists at least one shortest path from any member of that subset to other members [26]. The component ratio is defined as the number of components to the number of nodes. When $c$ denotes the number of components, the component ratio is $(c-1)/(n-1)$; where $n$ is the number of nodes.

## 4. Data Exploration

In the following, we explore the distinguishing power of the aforementioned measures in classifying malware families. We use various measures that examine the entire structure, substructure, and node properties in the social network analysis with their interpretation and how this interpretation is analogous to the problem at hand.

In order to estimate the behavior pattern of malware with respect to the influence of system calls, we measure the degree distribution, degree centrality, betweenness centrality, clustering coefficient, average distance, network density, and component ratio as features. Based on data exploration, we determine metrics for malware analysis. For data exploration, we used 3,615 malware samples representing 3 malware classes with 9 malware families and 153 benign samples; we will explain malware and benign samples in more detail in Section 6. We develop the malicious system call dictionary through empirical experiments with the various samples we had. We choose the system calls so that our system can detect and classify malware families within each malware class. We build the system call dictionary representing malicious activities, as outlined by the prior work on malware analysis [8, 29, 30]. To that end, we extract call sequence list of the system calls found in malware and their one-hop neighbors by comparing the call sequence of each malware sample with the predefined malicious system call dictionary. We transform such call sequence list into a system call graph to examine the social network properties. The details of the malicious system call dictionary are listed in the appendix.

*Degree Distribution*. First, we explore the degree distribution of malware system calls and outline various results. First, we observe that system calls with high in-degree centrality are the system calls found in malware related to creating local procedure calls (e.g., *NtAlpcSendWaitReceivePort* and *NtAlpcConnectPort* in adware) and synchronization (e.g., *NtReleaseMutant* in adware). On the other hand, we observe that system calls with in-degree centrality of 1 are mostly composed of system calls in the benign program related to file creation or opening, section creation, and process creation or termination; a few of system calls found in malware related to process creation (e.g., *NtCreateWorkerFactory* in adware) and synchronization (e.g., *NtCreateKeyedEvent* in adware) are included and exhibit the same feature. The details of the distribution of in-degree centrality in adware are listed in the appendix. Since system call patterns determine the program behavior, the malware calls out more system calls found in malware than system calls in benign program while conducting its malicious behavior. Thus, if the system calls with higher degree—the system calls found in malware—are deleted in the system call graph, those networks that exhibit the malicious behavior can collapse. We assume that system calls found in malware have significant influence on network structure of the system call graph such as degree distribution.

Figure 2 shows that the in-degree distribution of the malware system calls is highly right-skewed. The plot on the logarithmic scale in Figure 3 suggests that the in-degree distribution of the malware system calls follows a power-law distribution with noise in the right-hand side. To test the power-law hypothesis, we conduct a goodness of fit test over the binned empirical data [31]. We estimate the parameters $p(x_{min})$ and $\alpha$ of the power-law model as shown in Table 2. Since the $p$ value is larger than 0.1, the power-law is a plausible statistical hypothesis for our data set. Thus, we conclude that the degree distribution of malware system calls follows the power-law distribution approximately.

This test implies that a system call does not connect other system calls uniformly. Most system calls are connected to one system call and a few (popular) system calls are connected to numerous system calls. The portion of those calls with degree of one is significantly different among malware families as illustrated in Figure 2; for malware family the portion of calls with degree one is different from a

TABLE 2: The result of goodness of fit test.

| Malware ID[a] | $p(x_{min})$ | $\alpha$ | $p$ value |
|---|---|---|---|
| Win32.FakeInstaller 001 | .13889 | 2.21598 | .65634 |
| Win32.ScreenSaver 001 | .11765 | 1.86510 | .54446 |
| Win32.SideTab 001 | .04167 | 1.82649 | .17682 |

Malware ID[a]: the malware we used for experiments has a unique identifier as hash digest. For shortening the unique identifier, we used Malware ID as a sequence number.

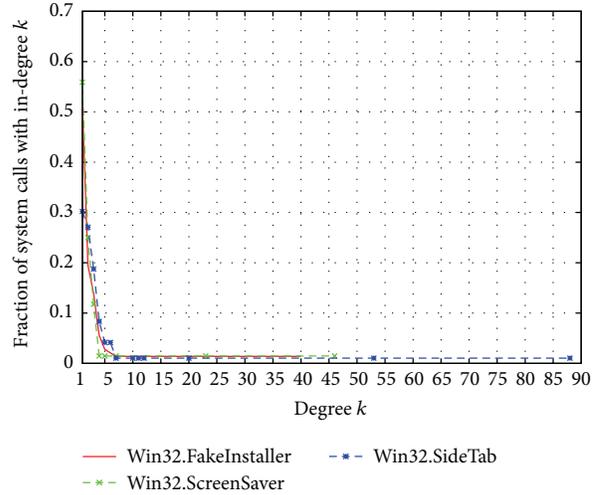

FIGURE 2: In-degree distribution of malware system calls (adware case).

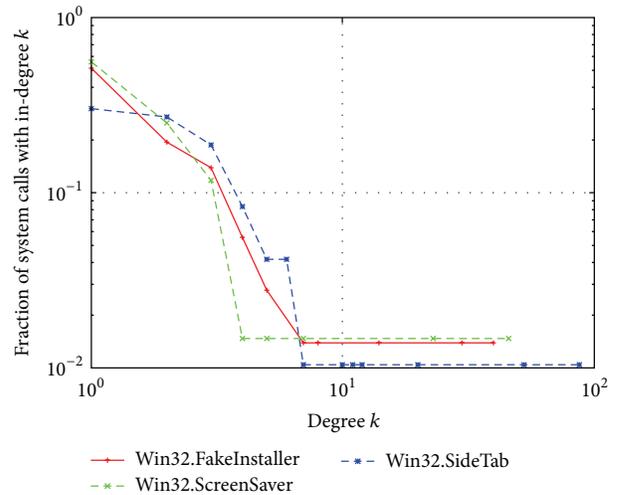

FIGURE 3: Log-scale in-degree distribution of system calls (adware case).

uniform distribution. The portion of system calls with degree of one gives hints on a good feature for classifying malware families.

Networks with power-law degree distribution are called scale-free networks, where they exhibit a number of interesting network properties with respect to degree distribution, centrality, and cohesion. We use network properties of the system call graph to classify malware. Based on such property, the portion of system calls with in-degree of one is much higher than that of system calls with in-degree greater than one, and the ratio varies according to the malware family in question. Thus, the difference in the portion of system calls with in-degree of one can be used as a major metric for classifying malware. The out-degree distribution of malware system calls is similar to the in-degree distribution and can be used—with similar rationale—as a distinguishing feature of malware families.

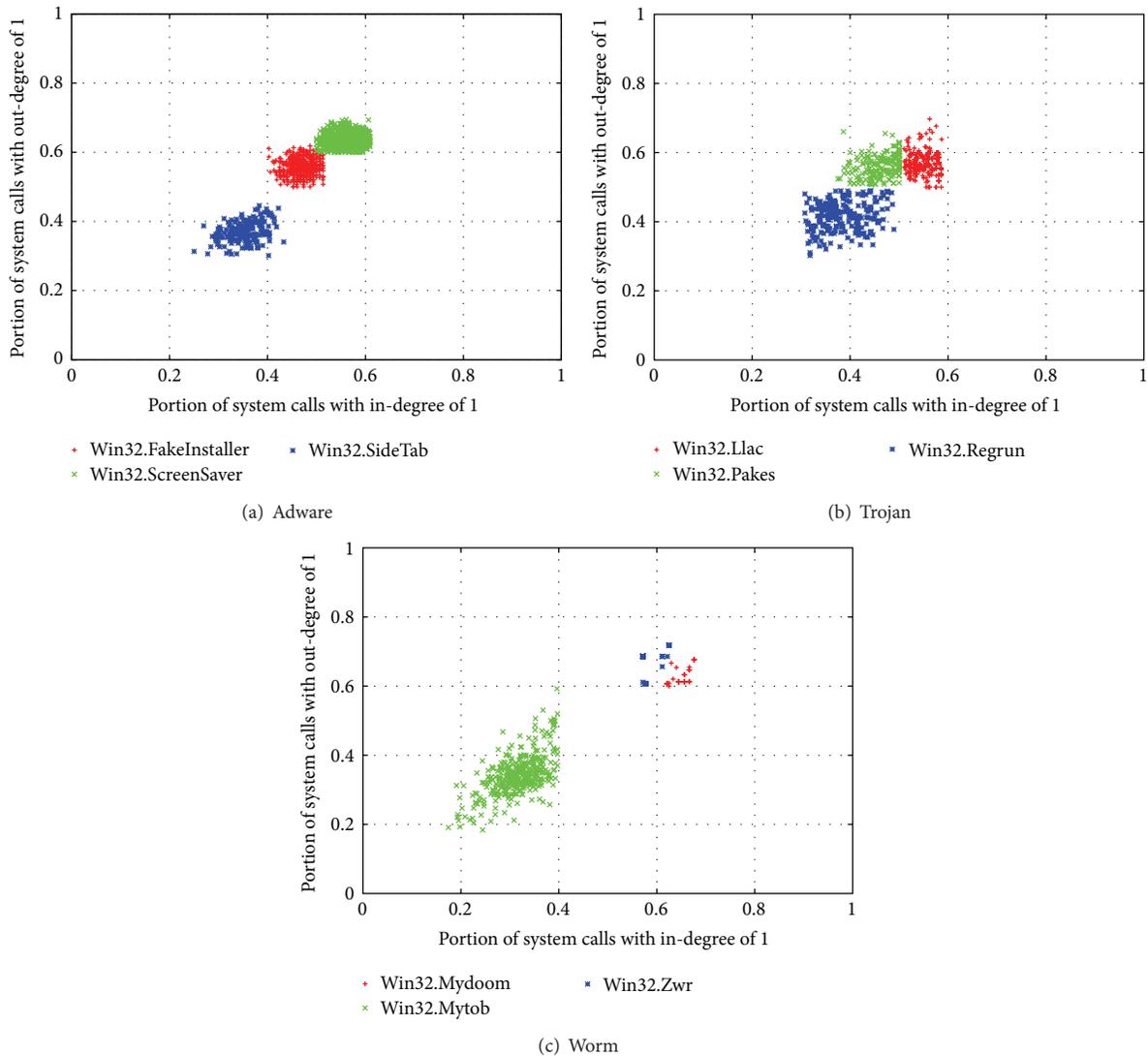

(a) Adware

(b) Trojan

(c) Worm

FIGURE 4: Distribution of portion of system calls for each malware family, (a) adware, (b) Trojan, and (c) worm (where in-degree and out-degree are 1).

*Road to Feature Selection.* Figure 4 shows that malware samples are grouped according to their family utilizing two portions of the system calls: the portion with in-degree of 1 and the portion with out-degree of 1. We thus choose these portions as principal features to classify malware. Note that the plot for weighted in- and out-degrees of 1 has approximately the same shape as that for in- and out-degrees of 1 as illustrated in Figure 4.

*Degree Centrality.* Second, we explore the degree centrality of malware system calls. For each system call, the sum of in-degrees is equal to the sum of out-degrees [26]; that is, the average degree centrality is twice the average in-degree centrality or the average out-degree centrality. For dimensionality reduction, we only consider the average in-degree centrality. The average in-degree centrality of system calls classifies each adware family—the results of data exploration for Trojan and worm are described in the appendix—as illustrated in Figure 5. In conclusion, we observe that the average in-degree centrality is also *good as a feature* for classifying Trojan and worm.

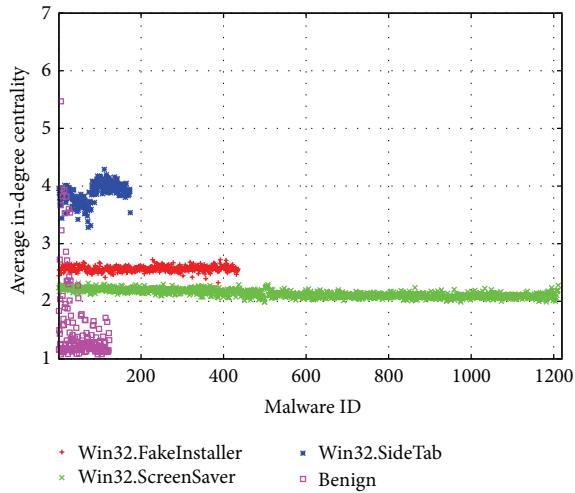

Figure 5: Average in-degree centrality (adware case).

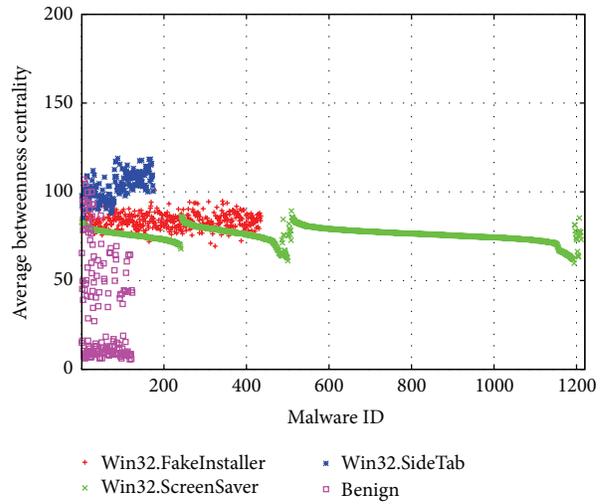

Figure 6: Average betweenness centrality (adware case).

*Betweenness Centrality*. Third, we examine the average betweenness centrality of each malware system call. We plot the average betweenness centrality of system calls according to adware sample in Figure 6. This shows that the average betweenness centrality of about half of Win32.ScreenSaver samples is similar to that of Win32.FakeInstaller samples. We also observe that the average betweenness centrality is a poor measure (compared to the previously examined measures) for classifying Trojan families. We notice that the average betweenness centrality is a good measure for only classifying worm families. Conceptually, centrality measures (e.g., betweenness centrality and degree centrality) represent a different process by which key nodes might influence the flow of information in a network. Measures of betweenness centrality and degree centrality are not occasionally highly correlated with each other, according to properties of networks [32]. Despite the general belief that such centrality is a good feature for classifying malware, in this study betweenness centrality is shown to be a poor feature for classifying malware. To this end, we consider that the average betweenness centrality is inappropriate for classifying malware in general when utilized alone.

*Structural Measures*. Fourth, we explore the overall structural information of the malware system call graph using the overall clustering coefficient, average distance, network density, and component ratio—all defined earlier. We examine the overall clustering coefficient of each malware system call, which is obtained by averaging the clustering coefficient over the number of system calls. We observe that the overall clustering coefficient is not a good basis for a classifier as illustrated in Figure 7, since many malware families such as Win32.SideTab and Win32.ScreenSaver overlap in such feature. We observe that the overall clustering coefficient is also a bad measure for classifying Trojan and worm families, although it could potentially be useful in classifying individual families. To this end, we conclude that the overall clustering coefficient is a bad measure and discard it as a potential feature.

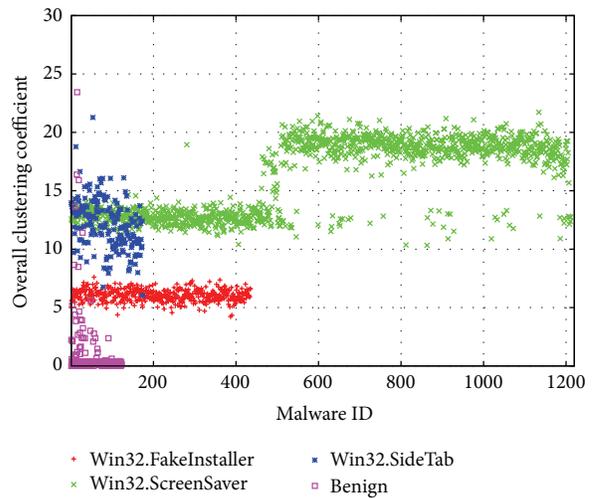

Figure 7: Overall clustering coefficient (adware case).

On the other hand, the average distance groups all adware samples according to each family as illustrated in Figure 8. We observe that the average distance is a good feature for classifying Trojan and worm families.

Next, we examine the network density and component ratio of each malware system call. We plot the network density of system calls according to adware sample in Figure 9, which shows that all malware samples—except the Win32.SideTab samples—have similar network density. To this end, we conclude that the network density is also not effective in classifying any of the Trojan and worm families.

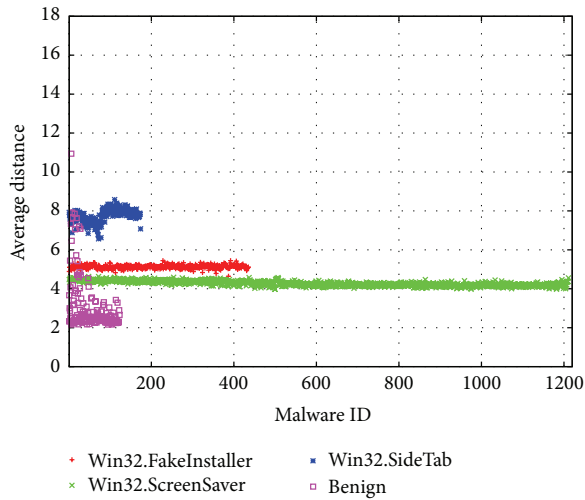

Figure 8: Average distance (adware case).

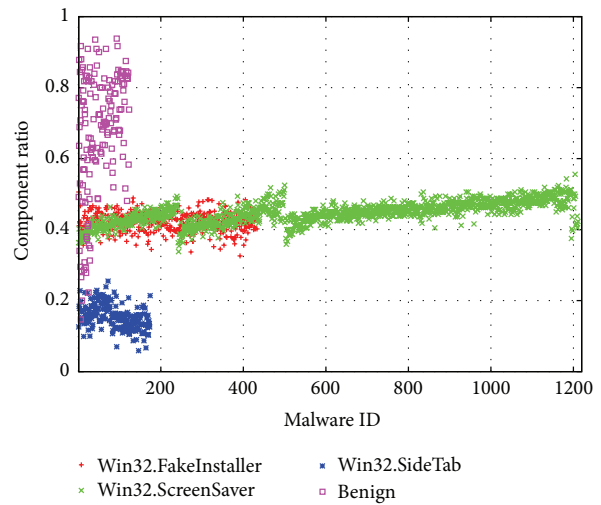

Figure 10: Component ratio (adware case).

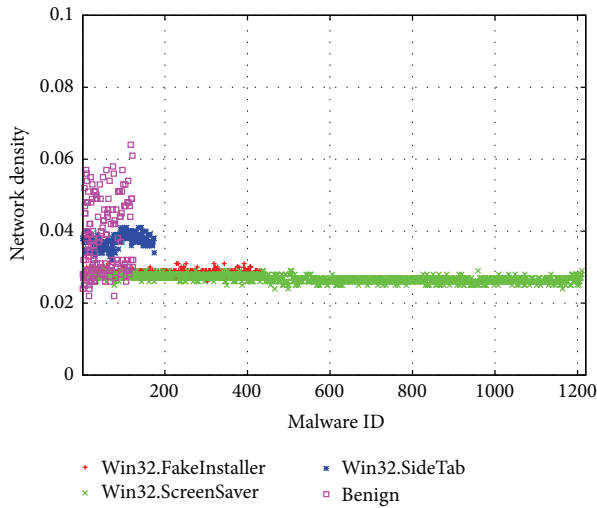

Figure 9: Network density (adware case).

Finally, we plot the component ratio according to adware sample in Figure 10, which shows that all malware samples—except the Win32.SideTab samples—have similar component ratio. In the case of Trojan, not all families can be classified using the component ratio. For example, the component ratio of about half of Win32.Zwr samples is similar to that of Win32.Mydoom samples with worms. Thus, the graph metrics of overall structure, such as the overall clustering coefficient, network density, and component ratio of system calls, are inappropriate for classifying malware. Notice that while the properties demonstrated in Figures 5–10 cannot be used individually to definitively classify malware, the partial power of these features can, when combined with other properties, assist the accuracy of classification.

Table 3 lists the results of additional data exploration that we conducted to find distinctive features for automatic classification. We find that the network diameter and H-index cannot be used for classification, whereas all other measures can be used to classify families in a binary learning process (two classes).

## 5. System Overview

After finding social network properties that distinguish behavioral traits of each malware sample, we implemented an automatic classification system based on such features. When a malicious program is passed to our system, it extracts system calls found in malware and their one-hop neighbors to examine their network (graph) properties (refined system call sequence). The system then compares the graph properties in the malicious program with those in each malware family and classifies malware to the proper families accordingly. As illustrated in Figure 11, our system is composed of three modules: the Behavior Identification Module, the Graph Metric Extraction Module, and the Classification Module.

*5.1. Behavior Identification Module (BI Module).* Our system analyzes malware through dynamic analysis. The BI module executes the given malicious program for 30 seconds and captures system call sequences through NtTrace (retrieved from http://www.howzatt.demon.co.uk/NtTrace/). The BI module then compares them with the predefined malicious system call dictionary, extracts system calls found in malware and their one-hop neighbors, and makes the refined system call sequences depict a system call graph.

*5.2. Graph Metric Extraction Module (Graph Module).* The graph module extracts the network properties according to the preselected metrics of SNA listed in Table 4. We notice that while the first set of the network-level properties are associated with what would seem to be a node-level feature, the fact that they are distributions of the feature makes them a network-level characteristic (as they are measured by the collective set of nodes).

Table 3: The result of additional SNA measure explorations.

| Category | Measure of SNA | Result of exploration |
|---|---|---|
| Centrality | Avg. normalized[a] betweenness centrality<br>Avg. normalized weighted[b] in-degree centrality<br>Avg. normalized weighted out-degree centrality | *Can classify only worm* |
| Cohesion | Network diameter, $H$-index | *Cannot classify all classes* |
|  | Number of strong connected components | *Cannot classify Trojan* |

normalized[a]: normalized betweenness/in-degree/out-degree centrality is the betweenness/in-degree/out-degree centrality divided by the maximum betweenness/in-degree/out-degree centrality.
weighted[b]: weighted degree centrality is the number of weighted links that a node has with other nodes in the network. The weighted links represent the amount of duplicated links between two nodes in the network.

Table 4: Measures of social network analysis for malware analysis.

| Scope | Category | Measure name |
|---|---|---|
| Node level | Centrality | Average in-degree centrality |
|  |  | Average weighted in-degree centrality |
| Network level | Degree distribution | Portion of calls with in-degree of 1 |
|  |  | Portion of calls with out-degree of 1 |
|  |  | Portion of calls with weighted in-degree of 1 |
|  |  | Portion of calls with weighted out-degree of 1 |
|  | Cohesion | Average distance |

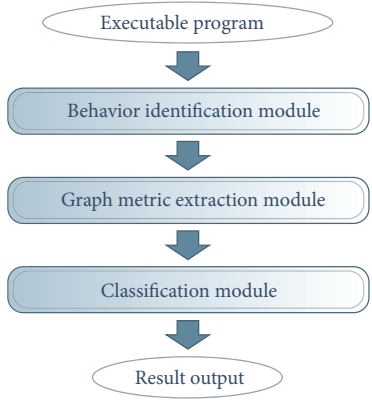

Figure 11: Overview of Mal-Netminer system.

Table 5: Malware classification and detection models using various learning algorithms.

| Category | Classifier | Previous works |
|---|---|---|
| Probabilistic-based algorithm | Naïve Bayes | [14, 19, 33–35] |
| Rule-based algorithm | RIPPER | [33, 35] |
| Function-based algorithm | SVM | [6, 7, 14, 34, 36–38] |
|  | MLP | [33, 38] |
|  | RBF | [39, 40] |
| Instance-based algorithm | $K$-NN | [14, 33, 34, 36, 38] |
| Tree-based algorithm | Decision tree | [6, 14, 16, 34, 36, 38] |
| Ensemble algorithm | Random forest | [38, 41] |
|  | Boosting | [14, 16] |
|  | Voting | [42] |
|  | Bagging | [41, 43] |

*5.3. Classification Module (Class Module).* There are various supervised learning algorithms in the machine learning community, including probabilistic-, rule-, function-, instance-, and tree-based algorithm and ensemble algorithms (based on the approach taken for solving the classification problem). Table 5 shows that previous work tried multiple classifiers for malware detection or classification. Following that, for an automatic classification, we choose Naïve Bayes, RIPPER, RBF, decision tree (C4.5), $K$-NN, and boosting (AdaBoost) classifiers. The class module utilized WEKA, a Java-based machine learning toolkit (retrieved from http://www.cs.waikato.ac.nz/ml/weka/).

*Naïve Bayes.* The Naïve Bayes classifier is a probabilistic classifier based on applying Bayes' theorem with independence assumptions [44]. For a given feature, this method computes the likelihood that a program is malicious. The method assumes that the attributes $X_1, X_2, \ldots, X_n$ are all conditionally independent of one another, for a given class label $y$. This dramatically simplifies the representation of $P(X_j \mid Y = y)$ and the problem of estimating it from the training data. The conditional independence assumption is given by

$$P(X \mid Y = y) = \prod_{i=1}^{n} P(X_j \mid Y = y). \quad (4)$$

*RIPPER.* RIPPER is a rule-based learning algorithm. It builds a set of rules that identify the classes while minimizing the amount of error. The error in RIPPER is determined by the number of training examples. This technique is easy to understand and is usually better than decision algorithm [45]. However, it is sensitive to noise and is not scalable as the training dataset size increases.

*Radial Basis Function (RBF) Network.* A Radial Basis Function (RBF) network is an artificial neural network that uses radial basis as activation functions; RBF is feed-forward networks trained using a supervised algorithm [40]. The naming of RBF comes from the fact that the basis functions in the hidden layer nodes are radially symmetric. The activation of a hidden layer node is computed by Gaussian function using the Euclidean distance or Mahalanobis distance between the input vector and the center of the basis function. The output of the network is a linear combination of radial basis functions of the inputs and neuron parameters. The neural network algorithms including RBF enable the performance of tasks which a linear program cannot solve but do not guarantee an optimal solution.

*Decision Tree (C4.5).* A decision tree is a decision support tool that uses a tree-like graph with internal nodes corresponding to attributes and leaf nodes corresponding to class labels [45]. Each internal node splits the instance space into two or more subspaces according to certain criteria such as information gain. This process is repeated recursively until a node has homogeneous characteristics. At that point, the class label is assigned. Most decision tree algorithms use the gain ratio to determine whether a certain internal node should be split or not. The merits of decision trees are that they are inexpensive to execute, fast at classifying unknown records, and easy to interpret when small enough.

*K-Nearest Neighbors (K-NN).* The $K$-NN algorithm is a type of lazy learning or instance-based algorithm. Lazy learning is a learning method in which the system delays generalization on the training data until it receives queries; in other words, the system simply stores training data and waits until it receives a test tuple [45]. The $K$-NN algorithm is one of the simplest algorithms in machine learning: a new instance is classified by a majority vote of its $K$ nearest neighbors. This technique needs much space to store the entire training data set and is sensitive to noise, but it effectively uses a richer hypothesis space in that it uses many local linear functions to form an implicit global approximation to the target function.

*Boosting (AdaBoost).* Boosting is one of the learning algorithms that combine multiple classifiers. It is a meta-algorithm that can be used in conjunction with many other learning algorithms to improve their performance. Boosting produces a set of weighted models by iteratively learning a model from a weighted data set, evaluating the model and reweighting the data set based on the performance evaluation results. AdaBoost (short for adaptive boosting) is an algorithm for constructing a strong classifier as a linear combination of simple weak classifiers through weighted vote. AdaBoost improves classification accuracy and generalizes fairly but does not guarantee an optimal solution. We used the AdaBoost algorithm to boost the Naïve Bayes, RIPPER, RBF, $K$-NN, and C4.5 algorithms.

## 6. Performance Evaluation

*6.1. Experiment Setup.* The 3,615 malware samples (our dataset is available at http://ocslab.hksecurity.net/mal_netminer) listed in Table 6 were collected through web crawling and malware repository such as Virusshare (retrieved from http://virusshare.com/) and Malwareblacklist (retrieved from http://malwareblacklist.com/); duplicated malware samples were eliminated based on the match of their SHA 256. We also excluded malware samples identified by fewer than 20 AV vendors included in the Virustotal dataset [46]. We used the textual description of malware that were produced by Kaspersky Lab (retrieved from http://www.kaspersky.com/). To get a representative set of benign samples, we used the 30 most popular programs (e.g., web browser and word processor) from [47] and 123 preinstalled benign programs in Windows system 32 directory; totally, we used 153 benign samples for our experiments. We performed all experiments in a hypervisor-based virtualization environment (VMware ESXi (retrieved from http://www.vmware.com/)); we performed all experiments on Intel Xeon X5660 and 4 GB of RAM with 32-bit Windows 7 Professional.

We implemented our proposed method with various parameters as follows:

(1) To extract the system call sequence list of Windows PE executables, the BI module was implemented as a python script coupled with NtTrace. In order to capture malicious behavior, the BI module executed the malware sample for 30 seconds. After that, our system converted the system calls log into gexf file for depicting call graph into dl file for the graph module consumption, successively.

(2) To examine the social network properties of each malware sample, the graph module utilized Ucinet 6 [48]. Since Ucinet 6 supports neither CLI (command line interface) nor API (application programming interface)—for automated analysis—the graph module used the GUI based hooking mechanism implemented in the python script. Our system stored the social network properties of each malware sample in a database.

(3) The "class module" utilized WEKA for the machine learning component of our system. We set up the training configuration values of WEKA through empirical experiments to achieve the best classification performance for each classifier as follows:

  (a) For Naïve Bayes, we modified the kernel estimator for numeric attributes.
  (to true)

  (b) For RIPPER, we modified the seed (to 13), the minimum total weight of the instance in a rule (to 1.5), and the number of optimization runs (to 10).

  (c) For RBF, we modified the seed to pass on to $K$-means (to 10), the maximum number of iterations for the logistic regression to perform

Table 6: Malware samples for experiments.

| Malware class | Malware family | Quantity | Behavior characteristics |
|---|---|---|---|
| Adware (1,820) | Win32.FakeInstaller | 435 | Download malicious files and display unwanted ads |
| | Win32.ScreenSaver | 1,211 | Display random ads in a pop-up |
| | Win32.SideTab | 174 | Download malicious files and display unwanted ads |
| Trojan (742) | Win32.Llac | 352 | Download files or give backdoor |
| | Win32.Pakes | 176 | Install "rogueware" by claiming the computer is infected by spyware |
| | Win32.Regrun | 214 | Perform various actions (e.g., spam emails and infect removable drives) |
| Worm (1,053) | Win32.Mydoom | 646 | Spread on the Internet as an attachment to infected messages |
| | Win32.Mytob | 358 | Similar to Mydoom, which provides backdoor for an attacker |
| | Win32.Zwr | 49 | Spread on WinRAR files by infecting RAR archives |

  (to 1), the minimum standard deviation for the cluster (to 0.07), and the number of clusters (to 4).
  (d) For C4.5, we modified the seed (to 10).
  (e) For $K$-NN, we modified the number of neighbors (to 3) and distance weighting method (weight by 1/distance).
  (f) For AdaBoost, we modified the seed (to 10) and the resampling (to true).

We used 5-fold cross-validation to evaluate the performance in our experiments. In the $k$-fold cross-validation, the whole data set is randomly partitioned into $k$ equal size subsamples; a single subsample is used as the validation data for testing the model, and the other $k-1$ subsamples are used as training data. We repeated the cross-validation process five times. All results were averaged over the five runs.

*6.2. Experiment Results and Analysis.* The performance evaluation focuses on the effectiveness of malware classification and discriminatory ability of the features. We used the accuracy as the performance metric, since the metric for performance evaluation must focus on the predictive capability of the model. We defined the accuracy as the entire number of the hits of the classifier (true positives + true negatives) divided by the whole dataset. The performance of malware classification model is determined by how well the model classifies various pieces of malware. Moreover, we used the receiver operating characteristic (ROC) curve as the metric for comparing classification models. The ROC curve is a plot of true positive rate (TPR) on the $y$-axis against false positive rate (FPR) on the $x$-axis. To compare the ROC performance of classifiers intuitively, we calculated the area under the ROC curve (the AUC) of each classifier, since the AUC represents the ROC performance in a single scalar value. If the area under the ROC curve (the AUC) is larger for one classifier than for others, that classifier is better and more robust.

*Comparing Different Methods.* To the best of our knowledge, the closest approaches in the literature to our approach are Fredrikson et al. [20] and Kolbitsch et al.'s work [8]. Fredrikson et al. [20] accurately distinguished unknown malware with an 86% detection rate, Kolbitsch et al.'s system [8] provided 93% of malware classification. For the completeness of our approach, we need to compare ours with these approaches. However, they are not available in public; we then conduct performance evaluation of ours by adopting various machine learning algorithms as the "class module."

*Discriminatory Ability between Malware and Benign.* When designing an antimalware system, one important factor that we should also consider is its discriminatory ability between malware and benign program. Antimalware systems must detect malware with small errors: false positive and false negative. Practically, however, we argue that it is more important for an antimalware system to detect malware with few small false negatives. On the other hand, one may consider an array of opposite opinions; for example, users can be bothered if their benign programs are misclassified as malware. To this end, we do not use those measures for performance evaluation to avoid the ambiguity of interpretation, because our method classifies three malware families and benign programs in each malware class. Accordingly, we used the accuracy and AUC as the performance metric.

Table 7 shows that the (boosted) RIPPER and boosted C4.5 classifiers outperform the other classifiers in terms of accuracy. The RIPPER classifier slightly outperforms the boosted C4.5 and boosted RIPPER classifier. The boosting algorithm somewhat improves the classification performance of all classifiers except for the $K$-NN and RIPPER classifier.

Several factors may have affected the experiment results. Since the social network properties of some Trojan samples, such as the average in-degree centrality and portion of system

Table 7: Comparison of classification accuracies and AUC for 3,615 types of malware and 153 benign samples.

| Classifier | Accuracy | | | AUC | | |
| --- | --- | --- | --- | --- | --- | --- |
| | Dataset_1[a] | Dataset_2[b] | Dataset_3[c] | Dataset_1 | Dataset_2 | Dataset_3 |
| Naïve Bayes | 99.96 | 90.19 | 95.16 | **1.00** | 0.95 | 0.98 |
| Boosted NB | 99.92 | 92.56 | 96.12 | **1.00** | 0.93 | 0.96 |
| RIPPER | 99.55 | **94.61** | 96.88 | 0.98 | 0.92 | 0.94 |
| Boosted RIPPER | 99.61 | 93.52 | 97.43 | **1.00** | **0.96** | **0.99** |
| RBF | 99.86 | 90.88 | 92.95 | **1.00** | 0.90 | 0.96 |
| Boosted RBF | **99.97** | 92.96 | 96.47 | **1.00** | **0.96** | 0.98 |
| C4.5 | 99.39 | 92.87 | 96.78 | 0.98 | 0.87 | 0.95 |
| Boosted C4.5 | 99.71 | 93.32 | **97.58** | **1.00** | 0.95 | **0.99** |
| K-NN | 99.13 | 93.05 | 96.20 | **1.00** | **0.96** | 0.98 |
| Boosted K-NN | 99.46 | 92.09 | 96.25 | 0.99 | 0.95 | 0.97 |

Dataset_1[a]: adware + benign, Dataset_2[b]: Trojan + benign, and Dataset_3[c]: worm + benign.

calls with in-/out-degree of 1, are similar to those of benign samples, classifiers need the discriminatory ability between some Trojan and benign samples that are nonlinearly separable. With the K-NN classifier, the lack of general information influences the overall performance. It is important to find an adequate K parameter for the (boosted) K-NN classifier. The choice of K involves an important tradeoff. On the one hand, if K is too large, some of the neighbors used to make a prediction will no longer be similar to the target of the prediction, and this may bias the prediction. On the other hand, if K is too small, the classifier will not have enough information, so the training dataset will possibly be misclassified. Since we set K = 3 in our experiment, the K-NN classifier substantially overfitted the input data. With the Naïve Bayes classifier, the assumption of class conditional independence, which does not hold perfectly in our dataset, causes some loss of accuracy. Since the RBF classifier is limited in classifying data samples with a large Mahalanobis distance from RBF center, the RBF classifier misclassifies some benign samples as Trojans.

In our evaluation, the boosted RIPPER and boosted C4.5 classifier were found to outperform other classifiers in terms of accuracy and AUC. While the boosted C4.5 classifier performed slightly better than the boosted RIPPER classifier, the difference of the performance was included in the performance variation of each classifier. Thus, we conclude that the boosted C4.5 classifier is the best classifier for its effectiveness of malware classification and detection. When a classifier learns a training set, the classifier decides an optimal weight of feature attributes considering their relative importance. We study what feature attributes are relatively more significant in classifying malware. We measure the importance of feature attributes as the information gain, defined as follows (of attribute ranking shown in Table 8):

$$IG(C, A) = H(C) - H(C \mid A),$$
$$\text{where } H(C) = -\sum_{c \in C} p(c) \log_2 p(c),$$
$$H(C \mid A) = -\sum_{a \in A} p(a) \sum_{c \in C} p(c \mid a) \log_2 p(c \mid a). \quad (5)$$

The information gain, $IG(C, A)$, measures the amount of entropy decease on a given class $C$, when providing a feature attribute $A$. The decreasing amount of entropy reflects the additional information gained by adding feature $A$. In the formula, $H(C)$ and $H(C \mid A)$ represent the entropies of class $C$ before and after observing feature $A$, respectively.

Table 8 shows how much each feature attribute has an effect on detecting and classifying malware, when utilizing the information gain for attribute ranking. The rank of importance of feature attributes changes according to the dataset, but the top feature in any dataset remains within the top three features for any other given dataset. With the ranking, we see that feature attributes such as the average weighted in-degree centrality, average in-degree centrality, and portion of calls with in-degree of 1 have a significant influence on the accuracy of classifier.

*Performance Advantage.* The proposed method can classify the malware with fewer features than previous methods adopting call graph in malware detection and classification. Moreover, our proposed method only takes 20 seconds (on average) to classify malicious executable. The majority of that time is spent to measure social network properties from Ucinet 6; it takes on average 15 seconds for the graph module to analyze malware. Since Ucinet 6 supports neither CLI nor relevant API, more reduction and optimization of Graph analysis time are out of the scope of this work. Nonetheless, the metrics that measure the social network properties of system calls found in malware enable us to build a quick and simple detection system against malware.

## 7. Discussion

*Mal-Netminer against Evasion Attacks.* Malware authors create new malware variants embedding various detection circumvention techniques, such as obfuscation and packing. Moreover, early work in the anomaly detection community showed how they can be subject to a slew of evasions [49, 50]. One wonders how resilient our approach is to evasions (e.g., mimicry attacks). Perhaps such attacks would create disconnected subgraphs with very distinct SNA properties.

Table 8: The importance of feature attributes based on information gain.

| Feature attribute | Dataset_1[a] | Dataset_2[b] | Dataset_3[c] |
|---|---|---|---|
| Average weighted in-degree centrality | **1.450** | 0.997 | **1.393** |
| Average in-degree centrality | 1.441 | 0.974 | 1.384 |
| Portion of calls with in-degree of 1 | 1.371 | **1.275** | 1.380 |
| Portion of calls with out-degree of 1 | 1.360 | 0.946 | 1.369 |
| Portion of calls with weighted out-degree of 1 | 1.107 | 0.675 | 1.302 |
| Portion of calls with weighted in-degree of 1 | 1.068 | 0.744 | 1.337 |
| Average distance | 1.015 | 0.430 | 1.369 |

Dataset_1[a]: adware + benign, Dataset_2[b]: Trojan + benign, and Dataset_3[c]: worm + benign.

However, when building system call graph, our system leverages the refined system call sequence. Our system extracts the system calls found in malware and their one-hop neighbors for behavior identification by comparing invoked system calls with the predefined malicious system call dictionary. Despite the fact that malware authors can add useless system calls for evasion, these system calls which are not included in the predefined malicious system call dictionary do not have an influence on the results of our system. At least any two added system calls will have edges with a system call in the malicious system call dictionary. This does not significantly change the structure of our proposed system call graph. Malware authors might downgrade the results of our system by considering the system calls which are one-hop neighbors of system calls in the malicious system call dictionary; we think complexities of these techniques are high.

*Mal-Netminer against Complicated Features.* Given that modern malware embed a mixed number of features (e.g., Trojan with rootkit and botnet functionalities), our approach has limitation to classify those malware samples strictly; it is far more interesting for vendors to classify malware samples with mixed features to understand the threats they need to deal with. However, without understanding the characteristics of basic malware classes such as adware, Trojan, and worm, we believe that it is impossible to detect and classify the malware with more complicated and mixed features. By studying malware with mixed features and adopting our system, we enable reacting to those malware.

*Dynamic Analysis without Interaction between Human and PC.* Our approach analyzes a target program without interaction between human and PC: autonomous installation and execution. One thinks that most of the interesting behaviors are not observed, particularly in the benign programs, and that could skew the results. It might be reasonable to assume that, in the absence of evasive malware, the malicious software would execute its malicious payload as executed, while benign programs might require user interactions (e.g., GUI-based utilities). However, autonomous installation and execution is an unavoidable procedure for automation of dynamic analysis. Depending on the number of samples to be analyzed, we can adopt manual human interactions to analyze these samples instead of Behavior Identification Module and afterward conduct the automatic classification procedure.

Table 9: Distribution of in-degree centrality in adware (Win32.ScreenSaver_001 case).

| System call | In-degree centrality |
|---|---|
| NtClose | 46 |
| **NtAlpcSendWaitReceivePort** | 23 |
| **NtAlpcConnectPort** | 7 |
| **NtReleaseMutant** | 5 |
| NtAllocateVirtualMemory | 4 |
| NtDuplicateObject, NtAlertThread, NtOpenThreadToken, NtQueryValueKey, NtUnmapViewOfSection, NtOpenKeyEx, NtCreateTimer, and NtQueryInformationProcess | 3 |
| NtQueryKey, NtSetEvent, NtOpenProcessToken, NtCreateEvent, NtQueryVirtualMemory, NtTestAlert, NtCreateThreadEx, NtDeviceIoControlFile, NtOpenProcessTokenEx, NtAlpcDeleteSecurityContext, NtResumeThread, NtCreateFile, NtWaitForSingleObject, NtAlpcSetInformation, NtCreateMutant, NtWaitForMultipleObjects, and NtOpenKey | 2 |
| **NtCreateWorkerFactory, NtCreateKeyedEvent**, NtOpenProcess, NtAccessCheckByType, NtSetValueKey, NtOpenEvent, NtSetInformationFile, NtCreateKey, NtOpenSection, NtAccessCheck, NtSetInformationThread, NtMapViewOfSection, NtCreateIoCompletion, NtDelayExecution, NtWaitForKeyedEvent, NtGetMUIRegistryInfo, NtFreeVirtualMemory, NtWaitForWorkViaWorkerFactory, NtQuerySystemInformation, NtEnumerateKey, NtEnumerateValueKey, NtOpenFile, NtMapCMFModule, NtQuerySystemInformationEx, NtQueryDefaultLocale, NtRequestPort, NtRequestWaitReplyPort, NtQueryAttributesFile, NtConnectPort, NtProtectVirtualMemory, NtWorkerFactoryWorkerReady, NtNotifyChangeKey, NtCreateSection, NtQueryInformationFile, NtAlpcCreatePort, NtSetInformationProcess, NtSetTimer, and NtTraceControl | 1 |

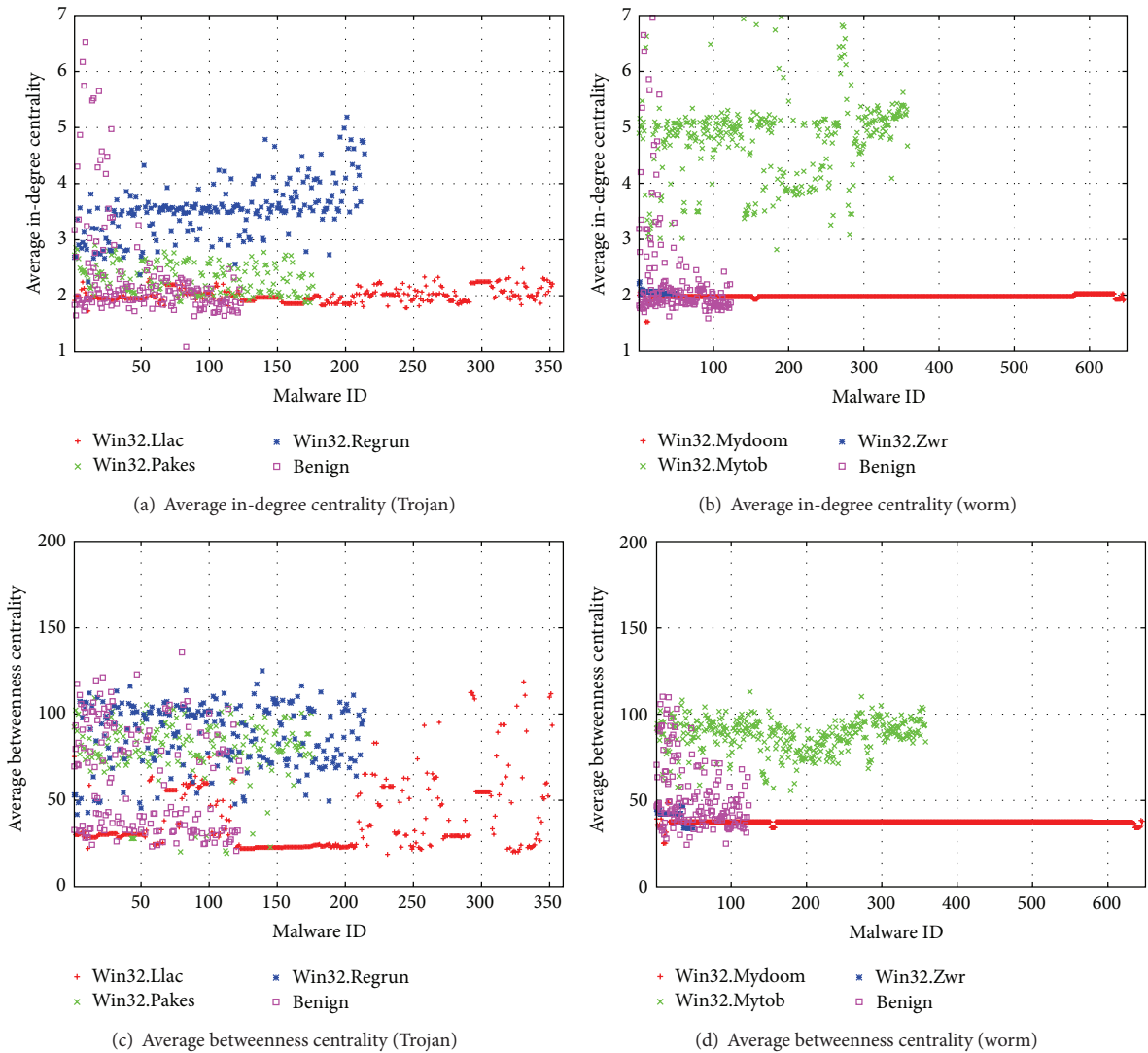

Figure 12: An evaluation of the node-level metrics on both Trojan and worm classes using the in-degree and betweenness centralities.

## 8. Conclusion and Future Work

In this paper, we presented a novel malware classification method that exploits social network properties as applied to call graphs from the dynamic execution of malware samples. Our proposed method is the first attempt to examine the topological structure of system call graphs and the node properties of system calls in malware. We explored graph metrics that measure the influence of system calls found in malware and those that measure the overall structure of malware. Through experiments, we found that the graph metrics of influence are effective for classifying malware, whereas the general structural information of malware, captured in the clustering coefficient, network density, and component ratio, is not effective for classifying malware.

Using this insight, we implemented a system for automatic malware classification. Our system extracts the system calls found in malware and their one-hop neighbors for behavior identification by comparing invoked system calls with the predefined malicious system call dictionary: making the refined system call sequence to depict a system call graph. It then measures graph properties and uses machine learning algorithms to train multiple classifiers on a set of malicious executables to classify new malware. Our experiments demonstrate that the proposed method performs well in classifying malware families within each class and detecting malware from benign programs. Our system provides more than 96% of classification accuracy. Unlike other systems, our system relies on a small set of features, enabling scalability.

*Future Work.* One major limitation of our system is that it classifies malware into broader classes, but not specific ones given by the market (e.g., Zeus, ZeroAccess, etc.). To this end, we plan to explore additional features that can be used for

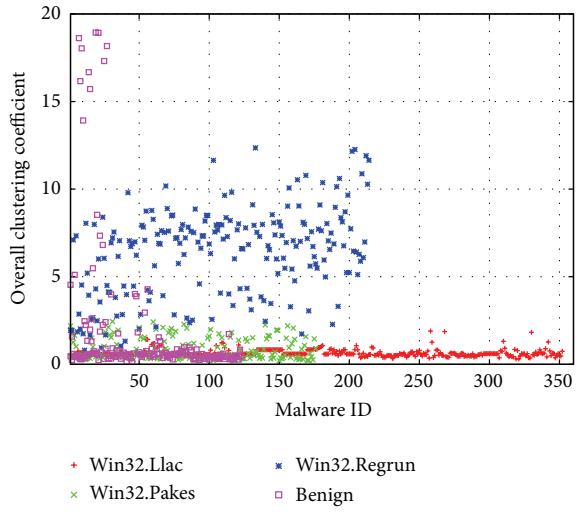

(a) Overall clustering coefficient (Trojan)

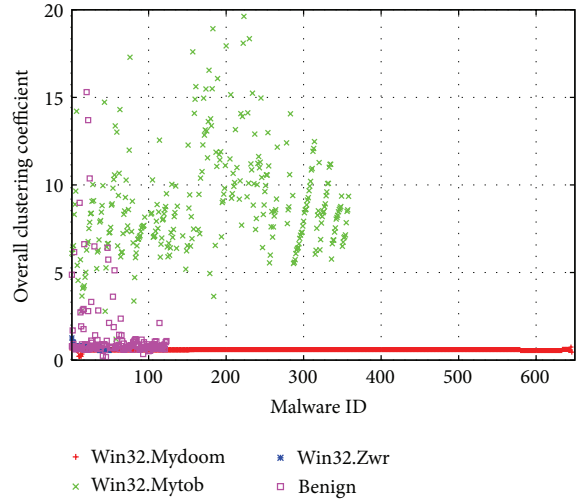

(b) Overall clustering coefficient (worm)

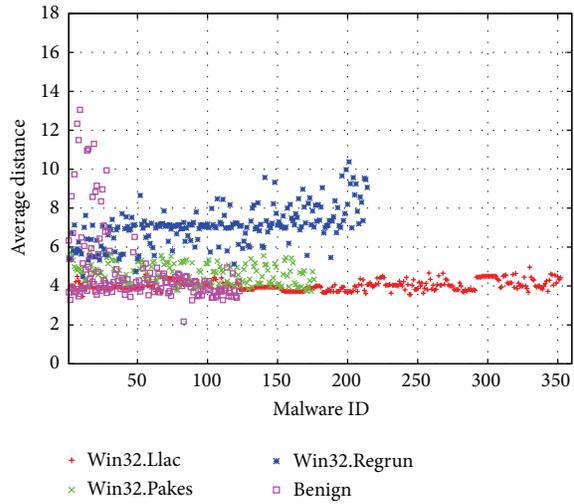

(c) Average distance (Trojan)

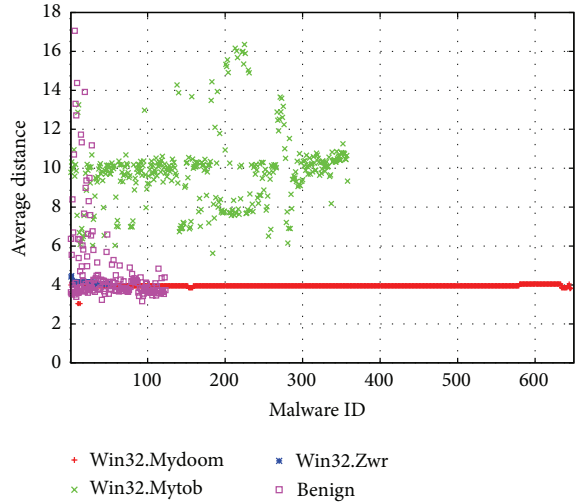

(d) Average distance (worm)

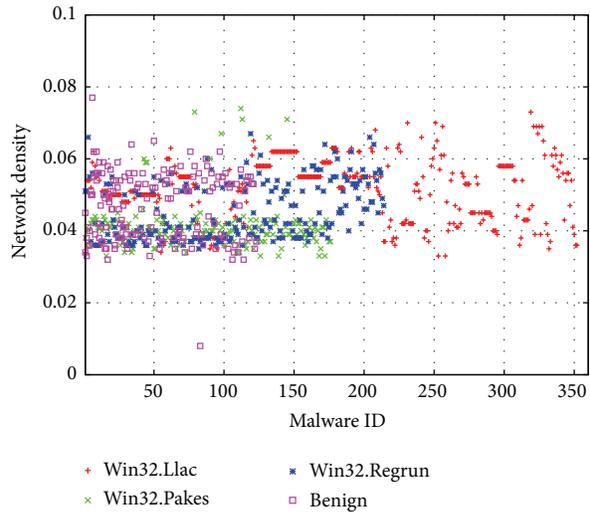

(e) Network density (Trojan)

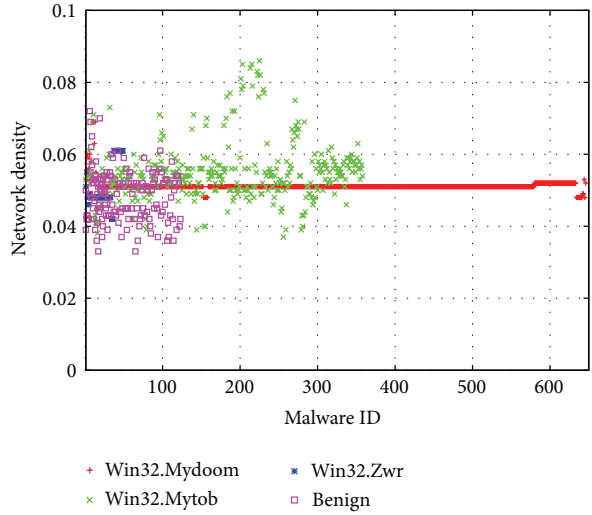

(f) Network density (worm)

Figure 13: Continued.

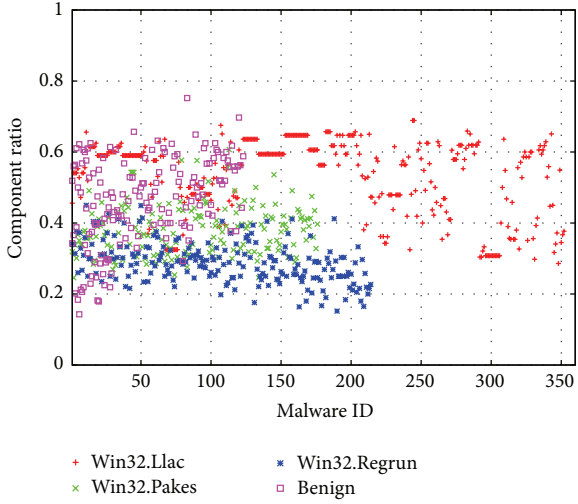
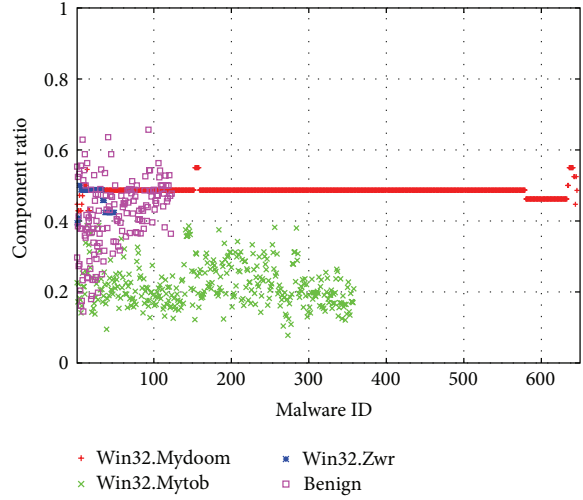

(g) Component ratio (Trojan)

(h) Component ratio (worm)

FIGURE 13: An evaluation of the network-level metrics on both Trojan and worm classes using the clustering coefficient, average distance, average density, and component ratio.

providing insight into lower granularity of malware classes. Malware clustering, using the features that we discussed in this paper, is yet another open direction.

## Appendices

### A. The Distribution of In-Degree Centrality in Adware

In the following, we outline the distribution of in-degree centrality in adware. In Table 9, we outline the distribution of in-degree centrality of Win32.ScreenSaver_001. Notice the frequency tendency of system calls invoked by the adware sample.

### B. The Detailed Malicious System Call Dictionary

In the following, we outline the detailed structure of the malicious system call directory. In Table 10, we outline the system call dictionary of adware. In Table 11 we outline the system call dictionary for Trojans. In Table 12 we show the system call dictionary of worms. Notice the differences between the different families and types of malicious pieces of malcode based on their dictionary structure and entries.

### C. Data Exploration for Trojan and Worm

*Metrics of Node Level.* For the node-level metrics studied in this paper, we choose the in-degree and betweenness centrality over the features studied in the paper. In particular and in line with the findings in the paper, the in-degree centrality for Trojan and worm is shown in Figures 12(a) and 12(b). Furthermore, the betweenness centrality for both classes is shown in Figures 12(c) and 12(d).

*Metrics of Network Level.* Network-level metrics include the overall clustering coefficient, average distance, network density, and component ratio. We complement those results shown in the paper with the following results. In Figures 13(a) and 13(b), we show the overall clustering coefficient for both the Trojan and worm classes. In Figures 13(c) and 13(d), we show the average distance for both Trojan and worm classes, respectively. In Figures 13(e) and 13(f), we show the network density for both the Trojan and worm classes, respectively. Finally, in Figures 13(g) and 13(h), we show the component ratio for the Trojan and worm classes, respectively. In all of these figures, findings are in line with the main findings in the paper, and all insight on those properties for the example class applies here as well.

TABLE 10: System call dictionary of adware.

| Functionality | System call list |
| --- | --- |
| Local procedure call | NtAlpcAcceptConnectPort, NtAlpcConnectPort, NtAlpcCreatePort, and NtAlpcSendWaitReceivePort |
| File & general I/O | NtCreateIoCompletion |
| Object | NtClose |
| Atoms | NtFindAtom |
| Processes & thread | NtResumeThread, NtCreateUserProcess, and NtCreateWorkerFactory |
| Synchronization | NtCreateKeyedEvent and NtReleaseMutant |
| Timers & system time | NtSetTimer and NtCreateTimer |

Table 11: System call dictionary of Trojans.

| Functionality | System call list |
|---|---|
| Processor & bus | NtFlushInstructionCache |
| Local procedure call | NtConnectPort, NtRequestWaitReplyPort, NtAlpcConnectPort, and NtAlpcSendWaitReceivePort |
| Memory | NtMapViewOfSection |
| File & general I/O | NtCreateFile, NtQueryInformationFile, and NtCreateIoCompletion |
| Object | NtClose |
| Atoms | NtAddAtom |
| Processes & thread | NtCreateThread, NtResumeThread, NtCreateProcessEx, NtQuerySystemInformation, NtCreateWorkerFactory, and NtQueryInformationProcess |
| Synchronization | NtCreateKeyedEvent and NtCreateMutant |
| Timers & system time | NtCreateTimer |

Table 12: System call dictionary of worms.

| Functionality | System call list |
|---|---|
| Processor & bus | NtFlushInstructionCache |
| Local procedure call | NtAlpcCreateSecurityContext and NtAlpcSetInformation |
| Memory | NtMapViewOfSection |
| Registry | NtEnumerateKey and NtEnumerateValueKey |
| Miscellaneous | NtQuerySystemInformation |
| File & general I/O | NtCreateFile and NtDeviceIoControlFile |
| Object | NtClose |
| Atoms | NtAddAtom |
| Processes & thread | NtCreateThread, NtResumeThread, NtCreateProcessEx, and NtQueryInformationProcess |
| Synchronization | NtReleaseMutant |
| Timers & system time | NtSetTimer and NtQueryPerformanceCounter |

## Conflict of Interests

The authors declare that there is no conflict of interests regarding the publication of this paper.

## Acknowledgments

This research was supported by the MSIP (Ministry of Science, ICT and Future Planning), Korea, under the ITRC (Information Technology Research Center) support program (IITP-2015-H8501-15-1003) supervised by the IITP (Institute for Information & communications Technology Promotion). In addition, this work was also supported by the ICT R&D Program of MSIP/IITP [14-912-06-002, The Development of Script-Based Cyber Attack Protection Technology].

## References

[1] AV-TEST: The Independent IT-Security Institute, 2015, http://www.av-test.org/en/statistics/malware/.

[2] KrebsonSecurity: Antivirus is Dead: Long Live Antivirus!, 2014, http://krebsonsecurity.com/2014/05/antivirus-is-dead-long-live-antivirus/.

[3] U. Bayer, P. Comparetti, C. Hlauschek, C. Kruegel, and E. Kirda, "Scalable, behavior-based malware clustering," in *Proceedings of the 16th Annual Network and Distributed System Security Symposium*, San Diego, Calif, USA, February 2009.

[4] U. Bayer, C. Kruegel, and E. Kirda, "TTAnalyze: a tool for analyzing malware," in *Proceedings of the 15th Annual Conference of the European Institute for Computer Antivirus Research*, pp. 180–192, 2006.

[5] D. Yuxin, Y. Xuebing, Z. Di, D. Li, and A. Zhanchao, "Feature representation and selection in malicious code detection methods based on static system calls," *Computers & Security*, vol. 30, no. 6-7, pp. 514–524, 2011.

[6] J. Dai, R. Guha, and J. Lee, "Efficient virus detection using dynamic instruction sequences," *Journal of Computers*, vol. 4, no. 5, pp. 405–414, 2009.

[7] B. Anderson, D. Quist, J. Neil, C. Storlie, and T. Lane, "Graph-based malware detection using dynamic analysis," *Journal in Computer Virology*, vol. 7, no. 4, pp. 247–258, 2011.

[8] C. Kolbitsch, P. Comparetti, C. Kruegel, E. Kirda, X. Zhou, and X. Wang, "Effective and efficient malware detection at the end host," in *Proceedings of the 18th USENIX Security Symposium*, pp. 351–366, Montreal, Canada, August 2009.

[9] D. Bruschi, L. Martignoni, and M. Monga, "Detecting self-mutating malware using control-flow graph matching," in *Proceedings of the 3rd International Conference on Detection of Intrusions and Malware & Vulnerability Assessment*, pp. 129–143, 2006.

[10] X. Hu, T.-C. Chiueh, and K. G. Shin, "Large-scale malware indexing using function-call graphs," in *Proceedings of the 16th ACM Conference on Computer and Communications Security (CCS '09)*, pp. 611–620, November 2009.

[11] J. Lee, K. Jeong, and H. Lee, "Detecting metamorphic malwares using code graphs," in *Proceedings of the 25th Annual ACM Symposium on Applied Computing*, pp. 1970–1977, March 2010.

[12] T. Wüchner, M. Ochoa, and A. Pretschner, "Malware detection with quantitative data flow graphs," in *Proceedings of the 9th ACM Symposium on Information, Computer and Communications Security (ASIA CCS '14)*, pp. 271–282, June 2014.

[13] M. Christodorescu and S. Jha, "Static analysis of executables to detect malicious patterns," in *Proceedings of the 12th USENIX Security Symposium*, pp. 169–186, Washington, DC, USA, August 2003.

[14] J. Z. Kolter and M. A. Maloof, "Learning to detect malicious executables in the wild," in *Proceedings of the 10th ACM SIGKDD International Conference on Knowledge Discovery and Data Mining*, pp. 470–487, Seattle, Wash, USA, 2004.

[15] D. K. S. Reddy and A. K. Pujari, "N-gram analysis for computer virus detection," *Journal in Computer Virology*, vol. 2, no. 3, pp. 231–239, 2006.

[16] S. M. Tabish, M. Z. Shafiq, and M. Farooq, "Malware detection using statistical analysis of byte-level file content," in *Proceedings of the ACM SIGKDD Workshop on CyberSecurity and Intelligence Informatics*, pp. 23–31, June 2009.

[17] M. K. Shankarapani, S. Ramamoorthy, R. S. Movva, and S. Mukkamala, "Malware detection using assembly and API call sequences," *Journal in Computer Virology*, vol. 7, no. 2, pp. 107–119, 2011.

[18] A. H. Sung, J. Xu, P. Chavez, and S. Mukkamala, "Static analyzer of vicious executables (SAVE)," in *Proceedings of the 20th Annual Computer Security Applications Conference (ACSAC '04)*, pp. 326–334, IEEE Computer Society, Washington, DC, USA, December 2004.

[19] C. Wang, J. Pang, R. Zhao, W. Fu, and X. Liu, "Malware detection based on suspicious behavior identification," in *Proceedings of the 1st International Workshop on Education Technology and Computer Science*, pp. 198–202, March 2009.

[20] M. Fredrikson, S. Jha, M. Christodorescu, R. Sailer, and X. Yan, "Synthesizing near-optimal malware specifications from suspicious behaviors," in *Proceedings of the 31st IEEE Symposium on Security and Privacy (SP '10)*, pp. 45–60, May 2010.

[21] M. Egele, T. Scholte, E. Kirda, and C. Kruegel, "A survey on automated dynamic malware-analysis techniques and tools," *ACM Computing Surveys*, vol. 44, no. 2, article 6, 2012.

[22] G. Cheliotis, "Social network analysis," Tech. Rep., The Saylor Foundation, 2013.

[23] M. E. J. Newman, "The structure and function of complex networks," *SIAM Review*, vol. 45, no. 2, pp. 167–256, 2003.

[24] A. A. E. Elhadi, M. A. Maarof, and B. I. A. Barry, "Improving the detection of malware behaviour using simplified data dependent API call graph," *International Journal of Security and its Applications*, vol. 7, no. 5, pp. 29–42, 2013.

[25] M. E. J. Newman, "Power laws, Pareto distributions and Zipf's law," *Contemporary Physics*, vol. 46, no. 5, pp. 323–351, 2005.

[26] M. Newman, *Networks: An Introduction*, Oxford University Press, Oxford, UK, 2010.

[27] M. E. J. Newman, D. J. Watts, and S. H. Strogatz, "Random graph models of social networks," *Proceedings of the National Academy of Sciences of the United States of America*, vol. 99, no. 1, pp. 2566–2572, 2002.

[28] A. Rusinowska, R. Berghammer, H. De Swart, and M. Grabisch, "Social networks: prestige, centrality, and influence," in *Relational and Algebraic Methods in Computer Science*, vol. 6663 of *Lecture Notes in Computer Science*, pp. 22–39, Springer, Berlin, Germany, 2011.

[29] L. Martignoni, R. Paleari, and D. Bruschi, "A framework for behavior-based malware analysis in the cloud," in *Proceedings of the 5th International Conference on Information Systems Security*, pp. 178–192, 2009.

[30] X. Wang, Z. Li, N. Li, and J. Choi, "PRECIP: towards practical and retrofittable confidential information protection," in *Proceedings of the 15th Network and Distributed System Security Symposium*, 2008.

[31] Y. Virkar and A. Clauset, "Power-law distributions in binned empirical data," http://arxiv.org/abs/1208.3524.

[32] T. W. Valente, K. Coronges, C. Lakon, and E. Costenbader, "How correlated are network centrality measures?" *Connections*, vol. 28, no. 1, pp. 16–26, 2008.

[33] F. Ahmed, H. Hameed, M. Shafiq, and M. Farooq, "Using spatio-temporal information in API calls with machine learning algorithms for malware detection," in *Proceedings of the 2nd ACM Workshop on Security and Artificial Intelligence*, pp. 55–62, 2009.

[34] I. Firdausi, C. Lim, A. Erwin, and A. Nugroho, "Analysis of machine learning techniques used in behavior-based malware detection," in *Proceedings of the 2nd International Conference on Advances in Computing, Control and Telecommunication Technologies*, pp. 201–203, 2010.

[35] M. G. Schultz, E. Eskin, E. Zadok, and S. J. Stolfo, "Data mining methods for detection of new malicious executables," in *Proceedings of the IEEE Symposium on Security and Privacy*, pp. 38–49, May 2001.

[36] A. Mohaisen and O. Alrawi, "Unveiling zeus: automated classification of malware samples," in *Proceedings of the 22nd International Conference on World Wide Web (WWW '13)*, pp. 829–832, May 2013.

[37] K. Rieck, T. Holz, C. Willems, P. Düssel, and P. Laskov, "Learning and classification of malware behavior," in *Proceedings of the 5th International Conference on Detection of Intrusions and Malware, and Vulnerability Assessment*, pp. 108–125, Paris, France, July 2008.

[38] R. Tian, R. Islam, L. Batten, and S. Versteeg, "Differentiating malware from cleanware using behavioural analysis," in *Proceedings of the 5th International Conference on Malicious and Unwanted Software*, pp. 23–30, October 2010.

[39] T. Liu, X. Guan, Q. Zheng, K. Lu, Y. Song, and W. Zhang, "Prototype demonstration: Trojan detection and defense system," in *Proceedings of the 6th IEEE Consumer Communications and Networking Conference (CCNC '09)*, pp. 1–2, IEEE, Las Vegas, Nev, USA, January 2009.

[40] Y.-X. Meng, "The practice on using machine learning for network anomaly intrusion detection," in *Proceedings of the International Conference on Machine Learning and Cybernetics (ICMLC '11)*, pp. 576–581, IEEE, Guilin, China, July 2011.

[41] M. Siddiqui, M. Wang, and J. Lee, "Detecting internet worms using data mining techniques," *Journal of Systemics, Cybernetics and Informatics*, vol. 6, no. 6, pp. 48–53, 2009.

[42] E. Menahem, A. Shabtai, L. Rokach, and Y. Elovici, "Improving malware detection by applying multi-inducer ensemble," *Computational Statistics & Data Analysis*, vol. 53, no. 4, pp. 1483–1494, 2009.

[43] T. Dube, R. Raines, G. Peterson, K. Bauer, M. Grimaila, and S. Rogers, "Malware target recognition via static heuristics," *Computers & Security*, vol. 31, no. 1, pp. 137–147, 2012.

[44] Theodoridis, Sergios and Koutroumbas, Konstantinos: *Pattern Recognition*, Academic Press, 4th edition, 2008.

[45] P.-N. Tan, M. Steinbach, and V. Kumar, *Vipin: Introduction to Data Mining*, Addison-Wesley Longman, Boston, Mass, USA, 1st edition, 2005.

[46] Virustotal, Virustotal, free online virus and malware scan, 2015, https://www.virustotal.com/.

[47] Download.com, CNET Download.com: free software download, 2015, http://download.cnet.com/windows/.

[48] S. Borgatti, M. Everett, and L. Freeman, *Ucinet for Windows: Software for Social Network Analysis*, Analytic Technologies, Harvard, Mass, USA, 2002.

[49] C. Kruegel, E. Kirda, D. Mutz, W. Robertson, and G. Vigna, "Automating mimicry attacks using static binary analysis," in *Proceedings of the 14th Conference on USENIX Security*

*Symposium—Volume 14 (SSYM '05)*, p. 11, Berkeley, Calif, USA, August 2005.

[50] D. Wagner and P. Soto, "Mimicry attacks on host-based intrusion detection systems," in *Proceedings of the 9th ACM Conference on Computer and Communications Security (CCS '02)*, pp. 255–264, ACM, November 2002.